\documentclass[8.5pt,twocolumn]{article}
\pdfoutput=1
\usepackage[super,sort&compress,comma]{natbib} 
\usepackage{mhchem}
\usepackage{times,mathptmx}
\usepackage{sectsty}
\usepackage{balance} 

\usepackage{graphicx}
\usepackage{lastpage}
\usepackage[format=plain,justification=raggedright,singlelinecheck=false,font=small,labelfont=bf,labelsep=space]{caption} 
\usepackage{fancyhdr}

\usepackage{hyperref}
\usepackage{abstract}
\usepackage{amsmath}
\usepackage[usenames,dvipsnames]{color}
\usepackage[percent]{overpic}
\usepackage{multirow}

\newcommand{\sub}[1]{\ensuremath{_{\textrm{#1}}}} 
\newcommand{\super}[1]{\ensuremath{^{\textrm{#1}}}} 

\newcommand{\labeledconfig}[2]{\begin{overpic}[width=0.49\columnwidth]{#1}\put(85,65){\large #2}\end{overpic}}

\definecolor{Blue1}{rgb}{.337254901, .705882352, .913725490}
\definecolor{Orange1}{rgb}{.901960784, .623529411, 0}

\newcommand{\edited}[1]{{#1}}

\newcommand{\tablecite}[1]{\edited{~[\citenum{#1}]}}

\begin{document}

\renewcommand{\thefootnote}{\fnsymbol{footnote}}
\renewcommand\footnoterule{\vspace*{1pt}%
\hrule width 3.4in height 0.4pt \vspace*{5pt}} 
\setcounter{secnumdepth}{5}

\makeatletter 
\def\subsubsection{\@startsection{subsubsection}{3}{10pt}{-1.25ex plus -1ex minus -.1ex}{0ex plus 0ex}{\normalsize\bf}} 
\def\paragraph{\@startsection{paragraph}{4}{10pt}{-1.25ex plus -1ex minus -.1ex}{0ex plus 0ex}{\normalsize\textit}} 
\renewcommand\@biblabel[1]{#1}            
\renewcommand\@makefntext[1]%
{\noindent\makebox[0pt][r]{\@thefnmark\,}#1}
\makeatother 
\renewcommand{\figurename}{\small{Fig.}~}
\sectionfont{\large}
\subsectionfont{\normalsize} 

\setlength{\arrayrulewidth}{1pt}
\setlength{\columnsep}{6.5mm}
\setlength\bibsep{1pt}

\twocolumn[
  \begin{@twocolumnfalse}
\noindent\LARGE{\textbf{Partial oxidation of Step-Bound Water Leads to\\Anomalous pH Effects on Metal Electrode Step-Edges}} 
\vspace{0.6cm}

\noindent\large{\textbf{
	Kathleen Schwarz,$^{\ast}$\textit{$^{a}$}
	Bingjun Xu,\textit{$^{b}$}
	Yushan Yan,\textit{$^{b}$} and
	Ravishankar Sundararaman\textit{$^{c}$}
}}\vspace{0.5cm}

\noindent \normalsize{\textbf{Abstract:} The design of better heterogeneous catalysts for applications such as fuel cells and electrolyzers
 requires a mechanistic understanding of electrocatalytic reactions
 and the dependence of their activity on operating conditions such as pH.
 A satisfactory explanation for the unexpected pH dependence of electrochemical
 properties of platinum surfaces has so far remained elusive,
 with previous explanations resorting to complex co-adsorption
 of multiple species and resulting in limited predictive power.
 This knowledge gap suggests that the fundamental properties of these catalysts
 are not yet understood, limiting systematic improvement.
 Here, we analyze the change in charge and free energies upon
 adsorption using density-functional theory (DFT) to establish
 that water adsorbs on platinum step edges
 across a wide voltage range, including the double-layer region, with a loss
of approximately 0.2 electrons upon adsorption.
 We show how this as-yet unreported change in net surface charge due to this water explains the anomalous
 pH variations of the hydrogen underpotential deposition (H\sub{upd}) and the
 potentials of zero total charge (PZTC) observed in published experimental data.
This partial oxidation of water is not limited to platinum metal step edges, and we 
report the charge of the water on metal step edges of commonly used catalytic metals, including
copper, silver, iridium, and palladium, illustrating that this partial oxidation of water
 broadly influences the reactivity of metal electrodes.}
\vspace{0.5cm}
 \end{@twocolumnfalse}
  ]

\footnotetext{\textit{$^{a}$~National Institute of Standards and Technology, Materials Measurement Laboratory, 100 Bureau Dr, Gaithersburg, MD, 20899 (USA).  E-mail: kas4@nist.gov}}
\footnotetext{\textit{$^{b}$~Department of Chemical and Biomolecular Engineering, University of Delaware, 150 Academy Street, Newark, DE 19716 (USA). }}
\footnotetext{\textit{$^{c}$~Joint Center for Artificial Photosynthesis, California Institute of Technology, 1200 E California Blvd, Pasadena, CA 91125 (USA). }}

Reducing dependence on fossil fuels to address energy security
and climate change issues requires new energy sources, more efficient
methods of energy conversion, and new methods for energy storage.
Electrochemical solutions to these problems, such as
fuel cells\cite{Steele2001,Debe2012,Arico2005}, batteries and
super-capacitors, are particularly promising since they are cleaner
and more efficient than conventional alternatives such as combustion.
Despite considerable efforts to understand and improve fuel cell catalysts,
characterizing and fully explaining even basic catalytic reactions
in the simplest electrochemical systems is often quite challenging.

Specifically consider platinum, which is central to modern electrochemistry
from its industrial applications in fuel cells\cite{Steele2001,Debe2012,Arico2005}
to its academic relevance as a standard for catalytic activity.
The Pt(111) surface, in particular, can be prepared reproducibly,
exhibits predictable behavior, and forms the basis
for understanding platinum electrodes in general.
However, polycrystalline platinum, platinum nanoparticles and other faces
of platinum display significantly  different pH-dependent behavior than Pt(111) for many reactions
and processes,\cite{Daiser1983,VanderNiet2013,VanderNiet2010,Koper2011,Koper2015,Skelton2000,Grecea2004}
including the onset of H\sub{upd} (hydrogen underpotential deposition),\cite{Gisbert2010,Sheng2015,Ramaker2015}
the kinetics of the hydrogen evolution/oxidation reactions,\cite{Durst2014,Kunimatsu2007,Strmcnik2013}
and the potential of zero total charge.\cite{VanderNiet2013,Climent1997,Frumkin1975}
These deviations from ideal behavior, which can significantly alter the operation
of the catalyst,\cite{Koper2011,Frumkin1968} are not yet completely understood.

Attempts to explain these individual phenomena separately invoke complicated combinations
of adsorbates as a function of pH and potential\cite{VanderNiet2013,Koper2015,Kolb2014,Janik2015}
with proposed explanations almost exclusively presuming that
adsorbates with integer charges are most likely.
For instance, the Koper group\cite{VanderNiet2013}
identified hydroxide adsorption as a possible explanation for the anomalous pH shift for 
the H\sub{upd} peak in the voltammogram.   As
the authors acknowledge, full coverage of hydroxyl groups would result in a much larger pH
effect than experimentally observed. They briefly consider partial charge (in their case, of
 the adsorbed hydroxyl group), then dismiss this possibility as ``somewhat artificial looking'',
 instead suggesting that coverage of hydroxyl groups on step-edge surfaces change with pH and
 potential, and are thus responsible for the anomalous pH effect of the onset of H\sub{upd}. 

While the assumption of integral (or nearly integral) charge appears
to be reasonable for many adsorbates on the Pt(111) surface, there is
 no fundamental reason for adsorbate charge
to remain integral since the metal is an electron reservoir.
Unsurprisingly, experimental evidence for non-integer adsorbate charges
on metal surfaces dates back to as early as 1939.\cite{Esin1939}

Formally, the quantity of interest is related to the electrosorption valency,
defined as the partial derivative of the surface charge density of the metal
with respect to the surface concentration of adsorbed species,
at constant electrode potential.
(See Ref.~\citenum{Schmickler2014} for a detailed review.)
The electrosorption valency can be experimentally measured,
and intuitively corresponds to the amount of charge that flows
to or from the electrode when an adsorbate moves from solution to the surface.

Few attempts have been made to use \emph{ab initio} techniques
to calculate electrosorption valencies and related quantities,\cite{Schmickler2014,Fang2014}
primarily because this requires an arbitrary partitioning between 
adsorbate and electrode of the spatially-continuous electron density
in a fixed-charge \emph{ab initio} calculation, and leads to
results dependent on the model for charge partitioning.

Recent developments of fixed-potential DFT methods\cite{Letchworth-Weaver2012,Sundararaman2012}
make it possible now to directly calculate the electrosorption valency and related quantities.
These calculations employ a grand canonical ensemble of electrons
at a fixed chemical potential set by the electrode potential.
The average number of electrons changes to minimize the grand free energy
and is not restricted to integer values.
Exactly as in real electrochemical systems, the net charge of the adsorbate
and metal surface is locally compensated by an induced charge density in the electrolyte,
treated using continuum solvation methods with ionic screening.\cite{Letchworth-Weaver2012,Gunceler2013}
With fixed-potential DFT methods, the charge changes continuously
upon adsorption at fixed potential. The difference in calculated charge of
a surface with adsorbate, from those of the solvated surface and adsorbate separately,
is equal to the charge that flows to/from the electron reservoir
(electrode in experiment), which corresponds precisely to the electrosorption valency.

However, the electrosorption valency as found experimentally, is relative to the solvated 
surface without the adsorbate.  In this context, even if a solvent molecule is specifically bound to the 
electrode surface and changes the number of electrons at the surface, the electrosorption valency of this 
solvent is defined to be zero.  To separate the effect of the specifically bound solvent which is in this case water,  we find the difference in charge
 between the surface with the specifically adsorbed water, and with a
hypothetical solvent that does not specifically bind (a continuum dielectric).  Hence, we are able to separate the effect of the 
specific binding of water from the dielectric contributions of the electrolyte.

Here we directly compute these charge differences for a set of adsorbates,
and we find that non-integer charge associated with oxidatively adsorbed water
on the step edges of platinum over a wide voltage range provides
a simple, universal and fundamental explanation to the apparently
disparate non-ideal phenomena on platinum surfaces other than Pt(111).\cite{
Daiser1983,VanderNiet2013,VanderNiet2010,Koper2011,Koper2015,Skelton2000,Grecea2004,Gisbert2010,
Sheng2015,Ramaker2015,Durst2014,Kunimatsu2007,Strmcnik2013,Climent1997,Frumkin1975}
We show that moving beyond the assumption of integer charge leads to a more
succinct description of electrochemical reactions with greater predictive power.
Most importantly, recently developed {\em ab initio} methods\cite{Letchworth-Weaver2012,Sundararaman2012,Gunceler2013}
can easily predict the (possibly non-integer) adsorbate charges, as discussed above.

To take a specific example, consider the underpotential adsorption of protons
from solution to hydrogen on platinum surfaces (H\sub{upd}).
At equilibrium,
\begin{equation}
G\sub{Pt} + G\sub{H$^+$} + n G\sub{e$^-$} = G\sub{Pt--H},
\end{equation}
where $G\sub{Pt}$, $G\sub{Pt--H}$, $G\sub{e$^-$}$ and $G\sub{H$^+$}$
are the Gibbs free energies of the bare platinum electrode,
hydrogen adsorbed on the platinum electrode, electrons from the electrode
and protons in solution respectively, and $n$ is the number of electrons
transferred from the electrode upon adsorption.
With changing pH, the proton free energy shifts as
$\Delta G\sub{H$^+$} = -RT \Delta\ln$(H$^+$ activity),
whereas with changing electrode potential, the electron free
energy shifts as $\Delta G\sub{e$^-$} = -F \Delta E$, where $F$ is Faraday's
constant and $E$ is the electrode potential.
Therefore, the electrode potential for proton adsorption changes
with pH as $-n F \Delta E = RT \Delta\ln$(H$^+$ activity),
which corresponds to $\Delta E/\Delta$pH$ \approx -59$~mV/$n$
at room temperature.
On the ideal Pt(111) surface, each proton gains one electron
upon adsorption as neutral hydrogen and $n = 1$ yields
the observed $-59$~mV shift per (increasing) pH unit.
However, we argue that other Pt surfaces contain oxidatively adsorbed water
on the step edges that must be displaced during H\sub{upd},
which requires the transfer of $n > 1$ electrons from the electrode
and hence explains the experimentally observed shift of magnitude
smaller than $59$~mV per pH unit.

The presence of the oxidatively adsorbed water similarly affects
the electron count of several reactions on Pt surfaces other
than (111) and leads to the anomalous pH dependence of
reactivity and the potential of zero total charge.
Accounting for non-integer charges, we therefore easily explain
all these phenomena without invoking complicated combinations
of adsorbates that the conventional analysis with integer charges
requires.\cite{VanderNiet2013,Ramaker2015,Kolb2014}

The first part of the paper uses computational methods to establish
that water binds strongly to step edges on platinum surfaces and
is partially oxidized throughout the relevant voltage range.
The second part compares computational and experimental data
for the anomalous pH dependence of H\sub{upd}, and
quantitatively demonstrates how that is a direct consequence of
H adsorption requiring the displacement of oxidatively bound water.
The computational results predict a larger pH dependence of H\sub{upd}
with increasing ionic strength, which could be experimentally tested.
The third part provides evidence that oxidative adsorption of water
is a general phenomenon on step edges in several other metals.

\section*{Methods}

We perform \emph{ab initio} calculations at fixed electron
chemical potential\cite{Letchworth-Weaver2012} using the framework of
Joint Density Functional Theory (JDFT)\cite{Arias2000,Petrosyan2005}
as implemented in JDFTx.\cite{Sundararaman2012}
In contrast to conventional fixed-charge calculations,
this allows for ready comparison with voltammetric data.
The calculations utilize a continuum solvent model
(LinearPCM continuum solvation model\cite{Gunceler2013} unless otherwise indicated) with
ionic screening (1~mol/L of cations and anions unless otherwise indicated).
This allows us to treat charged slabs with meaningful total free energies
and absolute electron chemical potentials (relative to the
vacuum level) that quickly converge with simulation cell size,
as detailed elsewhere.\cite{Letchworth-Weaver2012}
We relate the absolute electron potential to electrode potential
using the absolute potential of the standard hydrogen electrode (SHE)
as 4.68~V below the vacuum level for the LinearPCM solvation model,\cite{Gunceler2013}
and 4.55~V for the nonlocal SaLSA solvation model,\cite{SaLSA} following previous work.\cite{CANDLE}

The fixed electron chemical potential calculations directly provide
the total number of electrons $N$ in and the grand free energy $G$
of each slab configuration at a specified chemical potential $\mu$.
The difference between $N$ for a slab with an adsorbate and the total $N$
for the corresponding bare slab at the same potential and the isolated adsorbate
is then exactly the change in electron number due to adsorption, $n$,
that determines if the process is Faradaic or non-Faradaic and
the magnitude of anomalous pH shifts, if any.
Unlike most recent computational work that
focuses on potential-independent surface binding energies or
surface state charges,\cite{Fang2014} our fixed chemical potential
approach directly relates to experimental voltammograms.

We compare stabilities of various adsorbed species as a function
of electrode potential (Figure~\ref{fig:AdsorbateStability})
by calculating the grand free energy $G$ at one electron
chemical potential $\mu$ (specifically at 0~V SHE)
and extrapolating linearly to nearby potentials $\mu'$
using $G' \approx G - (\mu' - \mu) N$.
Note that the linearity is valid for \emph{specific} adsorbate configurations.
With changing potential, the most stable adsorbate configuration can change
from one to the other, e.g. from H to H\sub{2}O adsorbed around 0.25~V SHE
in Figure~\ref{fig:AdsorbateStability}(a), and the overall system free energy
tracks the lowest of all configurations and is not assumed to be linear.

Above, $G$ includes vibrational Helmholtz free energy contributions
(zero-point energy, finite temperature internal energy
and entropy contributions) of the adsorbed configurations.
We neglect the vibrational contributions of the Pt atoms in the slab
since they do not change appreciably between different adsorbate configurations.
Free energies of liquid H\sub{2}O and gas-phase H\sub{2} at STP
are calculated by adding experimental gas phase entropies\cite{CRC-handbook}
to solvated and vacuum DFT calculations respectively that
include the vibrational zero-point energy.
For liquid water, the gas phase entropy conveniently accounts for
vibrational, rotational and translational entropy of water molecules,
while the solvation model in the DFT calculation accounts for the
change in Gibbs free energy between the gas and liquid phases.
This directly gives us $G\sub{H\sub{2}O}$ for calculating adsorption free energies of water.
For proton adsorption, we calculate $G\sub{H$^+$} = \frac{1}{2} G\sub{H\sub{2}(STP)}$ at pH = 0
and at the SHE potential using the equilibrium that defines the standard hydrogen electrode,
and use the Nernst equation to calculate it for a different pH.
For hydroxide adsorption, we use $G\sub{OH$^-$} = G\sub{H\sub{2}O} - G\sub{H$^+$}$ 
from the ionic equilibrium in water.

We performed density functional theory calculations with JDFTx,\cite{Sundararaman2012}
with the same methodology as detailed elsewhere,\cite{Schwarz2015}
with the Perdew-Burke-Ernzerhof (PBE) exchange-correlation functional,\cite{Perdew1996}
and GBRV ultrasoft pseudopotentials\cite{Garrity2014}
with a 20 Hartree kinetic energy cutoff for the plane wave basis.
For the free energies of Figure~\ref{fig:AdsorbateStability}, we
 included pair-potential DFT+D2 dispersion corrections\cite{Grimme2006}
with a $C_6$ coefficient of 7 J nm$^6$ mol$^{-1}$ for Pt
to correctly account for binding energies on its surface.\cite{Blonski2012}
We used a $6 \times 8$ Monkhorst-Pack $k$-point mesh along the periodic directions,
with a Fermi smearing of 0.01 Hartrees for Brillouin zone integration.
We used primitive unit cells with 5 layers for the
(111) and (100) surfaces, 4 for (533) and (553),
and 3 for the (110) surface, along with a minimum separation
of 15 Angstrom between periodic images of the slabs
(which is then completely isolated using Coulomb truncation\cite{Sundararaman2013}).
The bottom layer of each slab is constrained to the calculated bulk lattice structure
and the remaining geometry is optimized fully for each calculation.

\section*{Results and Discussion}

\begin{figure}
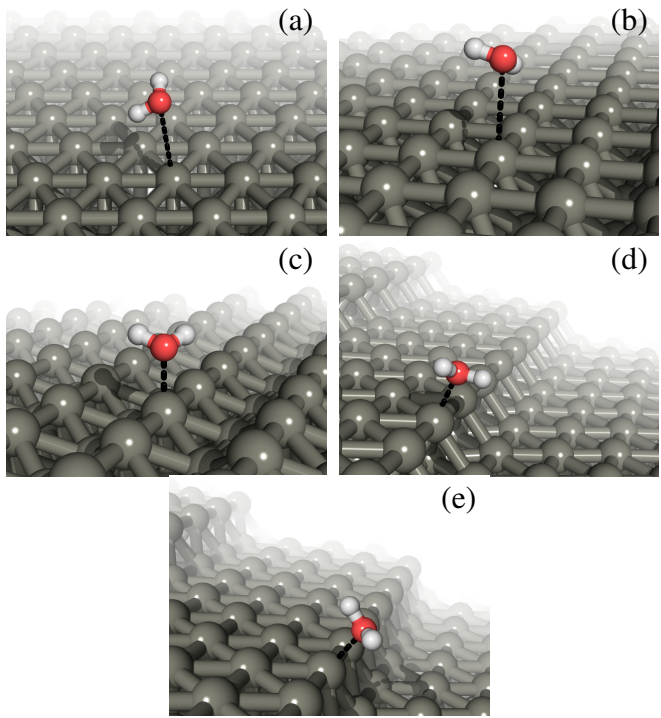

\centering{
\labeledconfig{111_Water}{(a)}\hfill\labeledconfig{100_Water}{(b)}\\
\labeledconfig{110_Water}{(c)}\hfill\labeledconfig{533_Water}{(d)}\\
\labeledconfig{553_Water}{(e)}
}
\caption{
Geometries of water molecules adsorbed at the atop
Pt sites on the five surfaces included in this study:
(a) (111), (b) (100), (c) (110), (d) (533) and (e) (553).
For clarity, we show water molecules from only one unit cell of the surface.
\label{fig:OneWaterConfigs}}
\end{figure}

We examine the properties of adsorbates on the five experimentally well-studied
Pt surfaces shown in Figure~\ref{fig:OneWaterConfigs}, which includes the low-index
(111), (100) and (110) surfaces as well as the stepped (533) and (553) surfaces.
The (533) surface has four-atom (111) terraces and (100)-like step edges, whereas the
the (553) surface has five-atom (111) terraces and (110)-like step edges.
These stepped surfaces serve as model surfaces for the more common
polycrystalline surfaces and nanoparticles which include similar step edges.

Determining the most stable adsorbate configuration is a difficult problem
because of the small energy differences that can be sensitive to the computational method.
Various studies differ on the most stable configurations of water, OH, H,
and famously CO\cite{Xu2014,Feibelman2001,Hayden1985,Lebedeva2002,Koper2000,Carrasco2012}
on various Pt surfaces, and even for the seemingly simplest case of H on Pt(111)\cite{Tan2013}.
Step edges are further complicated with chain-like water structures
reported in ultra-high vacuum experiments\cite{Morgenstern1996},
hexagonal and pentagonal arrangements predicted by DFT for partially-covered
step edges of Pt(533), and with energy differences less than 0.01~eV
for different configurations on fully-covered step edges\cite{WaterSolv2015}.
However, the adsorbate charges appear to be robust across various
low energy configurations (Table~\ref{tbl:AdsorptionCharges}) and therefore
we leave the exact determination of adsorbate configurations to a future study.
We consider the full-coverage adsorption of water to atop platinum sites
of step edges, which bind the strongest when available, or terrace sites
otherwise, as shown in Figure~\ref{fig:OneWaterConfigs}.

\begin{figure}
\centering
\includegraphics[width=\columnwidth]{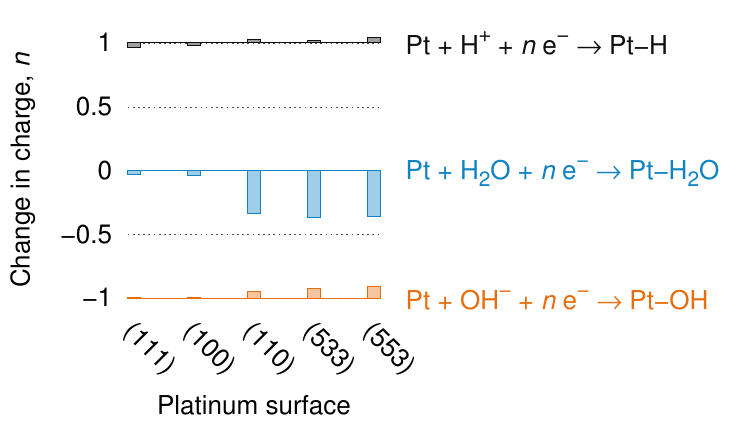}
\caption{Adsorption reactions (right) and corresponding change in number of electrons $n$ (left) for different Pt surfaces.
Lines indicate the conventionally expected integer charge, while the bars indicate the fractional deviations.
See Table~\ref{tbl:AdsorptionCharges} for the adsorption sites considered and the corresponding charges.
\label{fig:AdsorptionCharges}}
\end{figure}

\begin{table}
\centering{\begin{tabular}{cccc}
\hline\hline
Species & Surface & Site & $n$ \\
\hline
\multirow{5}{*}{OH${^-}$}
& 111 & atop & -0.99 \\
& 100 & atop & -1.00 \\
& 110 & atop & -0.94 \\
& 533 & atop & -0.92 \\
& 553 & atop & -0.91  \\
\hline
\multirow{8}{*}{H${^+}$}
& \multirow{3}*{111} & atop & 0.97 \\
&                    & fcc & 0.97 \\
&                    & hcp & 0.97 \\
& 100 & atop & 0.98 \\
& 110 & atop & 1.02  \\
& 533 & bridge  & 1.02  \\
& \multirow{2}{*}{553} & atop   & 1.04  \\
&                      & bridge & 1.02  \\
\hline
\multirow{5}{*}{Water}
& 111 & atop & -0.03    \\
& 100 & atop & -0.04      \\
& 110 & atop & -0.34          \\
& 533 & atop & -0.37     \\
& 553 & atop & -0.36     \\
\hline\hline
\end{tabular}}
\caption{Change in number of electrons, $n$, for H$^+$, OH$^-$ and H\sub{2}O 
adsorption on various sites of different Pt surfaces, relative to bare surface in dielectric solvent.
`Bridge' denotes bridging site between step edge platinum atoms.
For the stepped surfaces, `atop' sites refer to those on the step edge.
\label{tbl:AdsorptionCharges}}
\color{red}
\end{table}

For each of the surfaces of Figure~\ref{fig:OneWaterConfigs},
we plot the change in number of electrons upon adsorption of water molecules,
hydroxyl groups, and protons in Figure~\ref{fig:AdsorptionCharges},
with numerical results in Table~\ref{tbl:AdsorptionCharges}.
We report the charges calculated at 0~V relative to SHE, but these
charges change by less than 0.05 for electrode potentials between
 0~V and 1~V relative to SHE.
In all cases, hydroxides and protons transfer approximately one
electron to and from the surface respectively, as conventionally expected.
Water does not transfer a significant charge on Pt(111) or (100),
but adsorption atop the step-edge sites of the other surfaces
is accompanied by a loss of 0.3 to 0.4 electrons.
The step edge sites on the Pt(110) surface are strongly undercoordinated
which makes it favorable for them to strongly adsorb and partially oxidize water, and
the water binds closer to the Pt(110) step edge than it does to the terrace sites.  

\begin{figure}
\centering{\includegraphics[width=0.7\columnwidth]{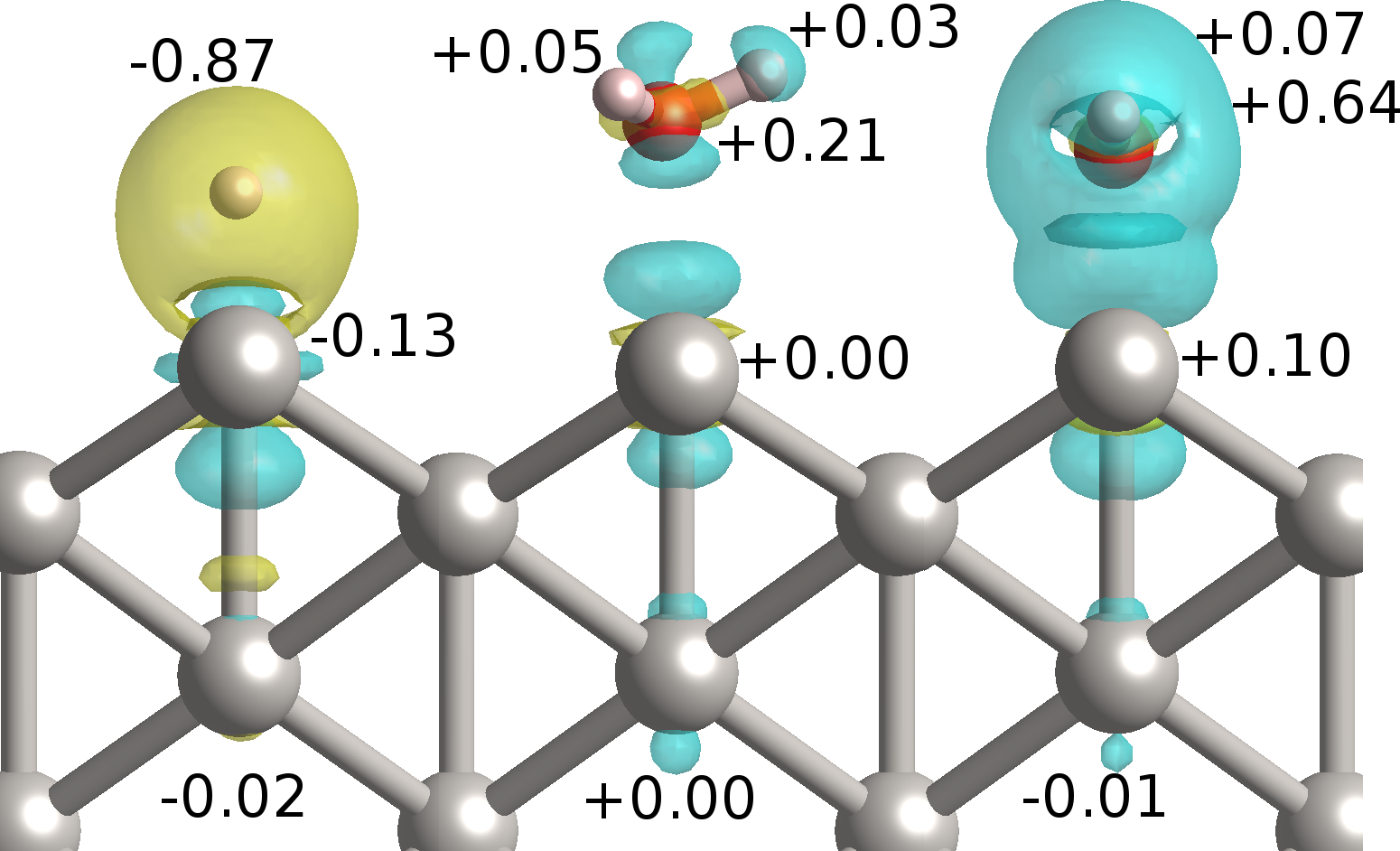}}
\caption{
Change in charge density upon 
adsorption of X = H$^+$, H\sub{2}O and OH$^-$ (from left to right) at the step 
edge of Pt(110), $\rho\sub{surf+X}(\vec{r}) - (\rho\sub{surf}(\vec{r}) + \rho\sub{X}(\vec{r}))$,
where the three $\rho(\vec{r})$'s in order are the charge density of the
adsorbed configuration, bare surface and isolated adsorbate species.
Cyan and yellow indicate the decrease and increase in electron density respectively.
The numbers indicate the corresponding changes in Lowdin charges upon adsorption.
\label{fig:AdsorptionChargeDensity}}
\end{figure}

The partial oxidation of water on the step edges makes performing fixed-potential
calculations over neutral, fixed-charge calculations essential.
In fixed-charge calculations of water adsorption on the step edge,\cite{WaterSolv2015}
when the cell is constrained to remain neutral, the water binds
much more weakly to the surface than the partially-oxidized water does.
In particular, we find that the neutral surface with adsorbed water adopts
a much lower potential (-1.03~V relative to SHE)
than the neutral bare surface (0.41~V relative to SHE),
so that the fixed-charge binding-energy is a difference of
energies at very different potentials and not relevant for electrochemistry.
Given that the water loses a charge equivalent to 0.34 electrons, the fixed-charge
(neutral cell) binding energy could therefore underestimate the more electrochemically meaningful
fixed-potential value by as much as $\sim (0.41-(-1.03))\times 0.34 \approx 0.5$~eV.
In contrast, this issue does not arise for adsorption on the terrace
because the water remains approximately neutral regardless.
Therefore we find a strong increase of 0.4~eV in the binding energy
of water on Pt(110) versus that on Pt(111) at 0 V vs SHE,
compared to that naively expected from neutral fixed-charge calculations
of potential-independent binding energies on the two surfaces.
This comparison highlights the importance of computing the electrosorption valency
of the adsorbates and performing fixed-potential calculations at the potential of interest.

Figure~\ref{fig:AdsorptionChargeDensity} shows the spatial distribution
of the electron density change upon H$^+$, water, and OH$^-$ adsorption at the Pt(110) step edge.
The reductive adsorption of H$^+$ leads to an increase in charge density, whereas
the oxidative adsorption of both the water and the OH$^-$ causes a loss in electron density
in the adsorbate and two Pt layers immediately below it.
Taking the difference in the Lowdin charges, we find that the charge assigned to
the oxygen increases by 0.21 electrons when the water adsorbs to the step edge.  
This accounts for most of the charge associated with the partial oxidation of the
water upon adsorption.  Interestingly, the charge associated with the platinum atom is nearly unchanged
by the presence of the water.  This suggests that the dipole moment of water is not the fundamental
reason for the partial oxidation.

Instead, the lone pairs on the oxygen atom on the water molecule appear
to play a role in the partial oxidation of the water.
The lone pairs are shared with the undercoordinated Pt step edge atoms
forming a stronger shorter bond, and in the fixed-potential scenario,
the excess electrons on the Pt are transferred to the reservoir
resulting in a net loss of electrons from the adsorbate + metal surface.
Similar behavior has been observed for weakly chemisorbed neutral adsorbates
with lone pairs on undercoordinated surface atoms, such as
pyridine on Au(210)\cite{gold210} and thiourea on mercury.\cite{pyridine}
In contrast, on the higher coordinated terrace sites, water adsorbs weakly
with mostly electrostatic interactions (dipole - induced dipole)
and little electron sharing, therefore resulting in negligible oxidation.

\begin{figure}
\includegraphics[width=0.495\columnwidth]{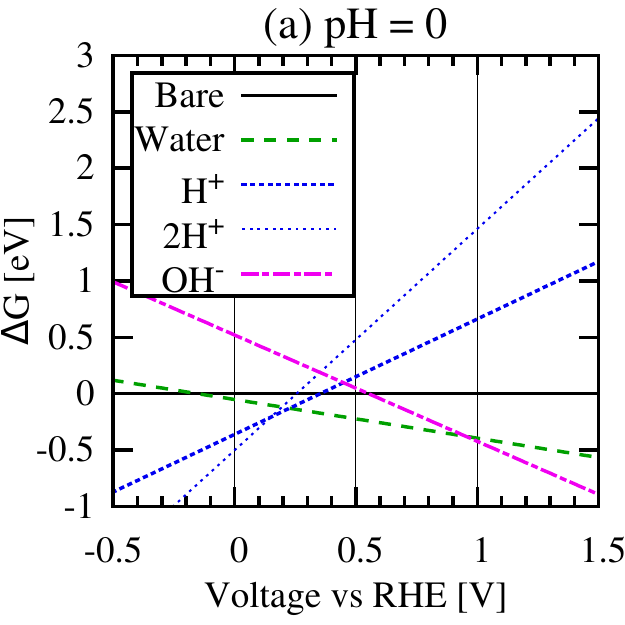}
\hfill
\includegraphics[width=0.495\columnwidth]{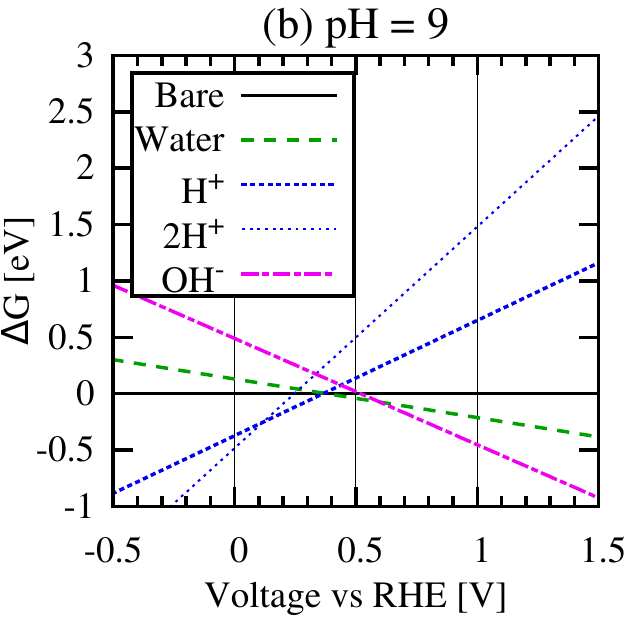}
\caption{
Gibbs free energies of H$^+$, water and OH$^-$ adsorbed on the
atop sites of the step edge of Pt(110), relative to the bare surface
in (a) acid conditions, pH = 0 and (b) basic conditions, pH = 9.
Free energies are estimated from calculations at 0~V vs SHE and include
vibrational and entropic contributions (see Methods section for details).
\label{fig:AdsorbateStability}}
\end{figure}

Next, we evaluate the prevalence of the oxidatively bound water on step edges
by calculating the relative free energies of water, protons and hydroxides
adsorbed on the Pt(110) surface over a wide voltage range at low and high pH.
Figure~\ref{fig:AdsorbateStability} shows that adsorbed water is stable
for a significant voltage range at pH = 0 that narrows with increasing pH.
 Further, water will be displaced by protons
at 0.25~V RHE for pH=0 which increases to 0.4~V RHE for pH = 9.

Our result for water adsorption resolves the long-standing debate regarding the species adsorbed
at the step edge, and it suggests that partially oxidized water likely plays 
a role in many of the reactions happening on the step edge, such as the 
hydrogen evolution reaction.\cite{Donadio2012,Durst2014,Kunimatsu2007}
Others have suggested that hydrogen binding energy\cite{Durst2014, Sheng2015}
changes with pH; this result illustrates that water oxidation rather than hydrogen
binding energy explains the unexpected pH dependence.

Figure~\ref{fig:AdsorbateStability} also illustrates that the
strongly bound, partially oxidized water must be displaced from
the step-edge during processes such as H\sub{upd} and OH adsorption.
This indicates that the H\sub{upd} reaction should be written as

\begin{equation}
y \textrm{H}^+ + \textrm{Pt--(H\sub{2}O})_x + n\textrm{e}^- \rightarrow \textrm{Pt--$y$H} + x \textrm{H\sub{2}O},
\end{equation}
where $y$ is the number of hydrogen atoms per Pt site, and $x$ is the number of displaced water molecules.

From this equation, the charge transferred per hydrogen atom during
H\sub{upd} will include both the charge from reduction of water as it is desorbed,
and the charge from reduction of H$^+$ as it is adsorbed.
Therefore hydrogen adsorption on platinum step-edges is not a one-electron process.
If $n$ electrons are transferred in H\sub{upd}, the electrode potential varies 
with pH as $E = E_0 + RT/(nF) \ln[\textrm{H}^+]$ by the Nernst equation,
where $F$ is the Faraday constant, $T$ is temperature, and $R$ is the gas constant.
At room temperature, $RT/F \ln[\textrm{H}^+] \approx -59$~mV/pH, 
so that $n > 1$ results in a reaction potential relative to
an absolute reference electrode that changes by less than 59 mV/pH. 
Relative to the RHE that shifts 59 mV/pH, this shift will be $59-59/n$~mV/pH.

\begin{figure}
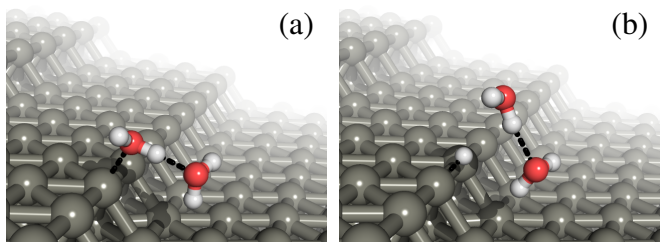

\labeledconfig{533_2Water}{(a)}\hfill
\labeledconfig{533_2WaterH}{(b)}
\caption{
Optimized structure of two water molecules per primitive cell of the
Pt(533) step edge, (a) before and (b) after a proton is placed on the step edge.
For clarity, we show adsorbates from only one unit cell of the fully-covered step edge.
\label{fig:TwoWaterConfigs}}
\end{figure}

\begin{table}
\centering{\begin{tabular}{ccc}
\hline\hline
\# waters & $n$ & Shift (mV/pH unit, relative to RHE) \\
\hline
1 & 1.39 & 17 \\
2 & 1.34 & 15 \\
\hline\hline
\end{tabular}}
\caption{
Change in number of electrons $n$ and corresponding anomalous potential shift 
per pH unit for the displacement of one or two water molecules by a proton.
\label{tbl:ChargeChanges}}
\end{table}

\begin{table}
\setlength{\tabcolsep}{2pt}
\centering{\begin{tabular}{c|cc|cc|cc}
\hline\hline
\multirow{3}{*}{System} & \multicolumn{6}{c}{Shifts [mV vs RHE / pH unit]}\\
\cline{2-7}\phantom{\LARGE A}
& \multicolumn{2}{c|}{Experiment} & \multicolumn{2}{c|}{SaLSA\cite{SaLSA}} & \multicolumn{2}{c}{LinearPCM\cite{Gunceler2013}}\\
& 0.1M [Ref] & 0.2M [Ref] & 0.1M & 1M & 0.1M & 1M\\
\hline
110 step & 10\tablecite{Sheng2015} & 11\tablecite{Gisbert2010} &&&&\\
(553) & & 10\tablecite{VanderNiet2013} & {14} & {16} & {15} & {17}\\
\hline
100 step & 8\tablecite{Sheng2015} & 11\tablecite{Gisbert2010} &&&&\\
(533) & & 10\tablecite{VanderNiet2013}  &  {14} & {16} & {15} & {17}\\
\hline
(110) & & & 11 & 13 & 14 & 16 \\
\hline\hline\end{tabular}
}
\caption{
Observed and predicted H\sub{upd} shifts for platinum surfaces and step edges.
The Pt(553) and Pt(533) surfaces have 110 and 100 step edges respectively; the rows
labeled `step' correspond to experiments on step-edge sites of nanoparticles.\cite{Sheng2015,Gisbert2010}
\label{tbl:HupdPZTC}}
\end{table}

\begin{figure}
\centering
\includegraphics[width=\columnwidth]{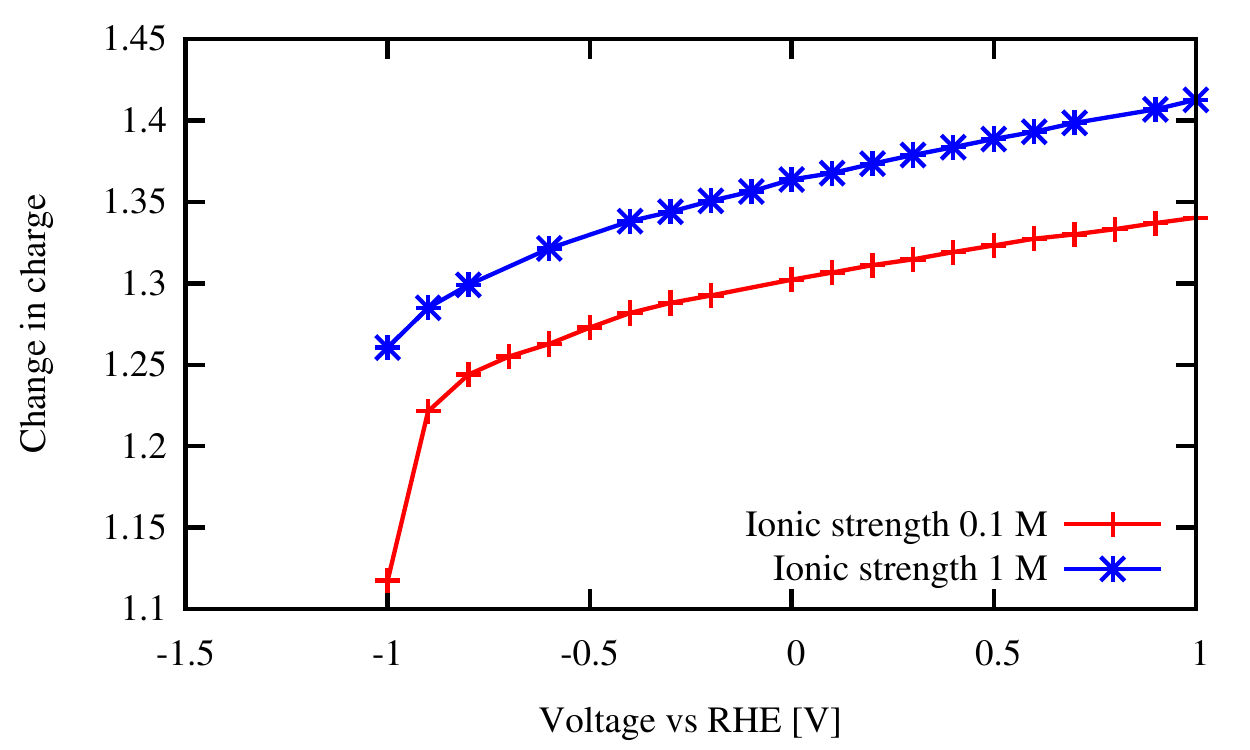}
\caption{Change in charge upon displacement of H\sub{2}O adsorbed on Pt(110) by H,
as a function of electrode potential, showing the effect of ionic strength.
\label{fig:Electrosorption}}
\end{figure}

\begin{figure}
\centering
\includegraphics[width=\columnwidth]{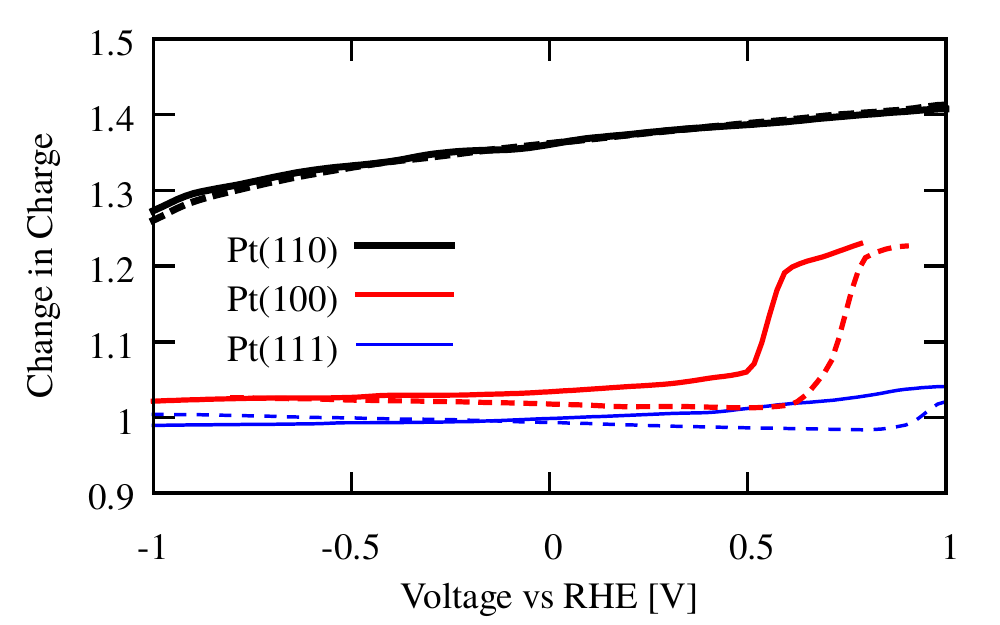}
\caption{Change in charge upon displacement of H\sub{2}O
adsorbed on low-index surfaces by H, as a function of electrode potential.
The solid lines include dispersion corrections, while the dashed lines do not.
Dispersion corrections negliblibly alter results for the strongly-bound water on (110),
but change the distance and consequently the charge changes for the weakly bound water on (100) and (110).}
\label{fig:LowIndexPlanesVoltage}
\end{figure}

Table~\ref{tbl:HupdPZTC} summarizes the predicted anomalous pH effect
from experimental data and from DFT calculations at 0 V vs SHE.
The experimental data range from 8~mV/pH to 11~mV/pH,
corresponding to $n=1.16$ to $n=1.23$, rather than
the expected 1 electron per proton.
The DFT calculations range from $n=1.25$ to $n=1.33$, corresponding to 
12~mV/pH to 15~mV/pH, with the 
nonlocal SaLSA model results falling closer to the experimental data range. 
The LinearPCM continuum model overestimates solvation of charged solutes
in general, and positively charged species in particular;\cite{SaLSA} 
therefore it overestimates the partial oxidation of water here as well.

Table~\ref{tbl:HupdPZTC} also illustrates the effect of ionic strength.  The ionic strengths used in experiment
vary, with the experiments for the 110 step edge with 0.1 M ionic strength\cite{Gisbert2010}
and 0.2 M ionic strength\cite{Sheng2015}.  Figure~\ref{fig:Electrosorption} illustrates that
increasing the ionic strength increases the expected value of $n$.
Increasing ionic strength allows for more charging on the surface,
which is known to increase the value of the electrosorption valency\cite{Guidelli1998}.  Additionally, one can see 
that the partial oxidation from water binding changes
to nearly zero at low potentials near the potential for water reduction.  

The change in charge upon displacement of H\sub{2}O by H is dramatically larger for the
Pt(110) surface than the other low-index surfaces, as shown in Figure~\ref{fig:LowIndexPlanesVoltage}.
This change in charge is nearly constant across a wide voltage range for the three surfaces, illustrating
the difference between the stepped surface of the Pt(110) and the other surfaces.  At high
enough potentials, the water begins to interact more strongly with the Pt(100) surface,
 but this happens at a potential well beyond that of the underpotential deposition of hydrogen.  We note that
the transition potentials here should only be qualitatively interpreted, given the limitations of solvation models.

To further ensure the reliability of our DFT results, we consider three possible complications:
 1) the possibility that the value of $y$, the number of hydrogen atoms per Pt site, is not one
2) the configuration of water molecules (and the possibility that $x$, the number of adsorbed water per Pt site, is not one)
 and neighbor effects between adsorbed water and H
3) accuracy of DFT or the solvation method

First, we consider the possibility that more than one proton is bound to a given Pt site.
DFT studies of Pt(553) surfaces show that hydrogen atoms adsorb
most favorably on the atop, fcc hollow, hcp hollow and bridge sites
at the step edge with very similar binding energies.\cite{Kolb2014}
Identifying the lowest free energy configuration of two protons requires
an expansive search,\cite{Tan2013} but the free energy will vary negligibly
between configurations due to the similarity in binding energies.
Here, we consider one possible configuration of two protons
atop the step edge platinum atoms of Pt(110).
Free energy calculations shown in Figure~\ref{fig:AdsorbateStability}
indicate that it is energetically favorable for a single proton
to displace water, rather than a two proton (`2H$^+$') configuration,
which only becomes energetically favorable at lower voltages.

Next, the configuration of the water molecules on Pt step edges
is likely to change with experimental conditions.
To assess how this affects the oxidation state of the adsorbed water,
we consider two water molecules per Pt site, with the water molecules occupying
both the top and bottom of the step edge as shown in
Figure~\ref{fig:TwoWaterConfigs}, which is similar to the favorable
adsorption structure identified by Kolb et al.\cite{WaterSolv2015}.
Note that we do not calculate with more water layers because they
do not have a single low energy configuration, which necessitates
\emph{ab initio} molecular dynamics with substantially
higher computational cost and complexity of analysis.
We then added H to the bridge sites, re-optimized the geometries
and find that H displaces the water from the step edge with a computed pH
shift almost identical to the previous case (Table~\ref{tbl:ChargeChanges}),
with the additional water resulting in a change in the pH shift of only 2~mV/pH unit.
Therefore, the charge difference is relatively insensitive to coverage
and nearest neighbor effects, and the inclusion of further solvent molecules
with molecular dynamics is unlikely to substantially alter our predictions.

Further, we examine the variation of the adsorption charge of water
on the step edge of Pt(110) with respect to DFT functional and solvation method, at 1 M ionic strength.
For the PBE exchange-correlation functional and the LinearPCM solvation model
from Ref.~\citenum{Gunceler2013} that we used for much of this paper,
the change in electron count is -0.34.
With the same solvation method, the local LDA functional and hybrid PBE0 functional
yield -0.35 and -0.37 respectively, whereas using the PBE DFT functional with the
NonlinearPCM solvation method from Ref.~\citenum{Gunceler2013} yields -0.36
and the nonlocal SaLSA solvation method from Ref.~\citenum{SaLSA} yields -0.26.
Therefore, the charges we predict are insensitive to the
DFT functional with a variation less than 0.02 electrons,
and only slightly sensitive to the solvation method.

This insensitivity in the change in charge over a wide voltage range
and over differing environmental conditions provides evidence of
the robustness of our result that the water oxidizes on step edges
throughout a large voltage range.
Having identified the source of the H\sub{upd} anomalous
pH effect, we can explain other effects such as
the anomalous potential of zero total charge (PZTC).
The partially oxidized water shifts the step-edge H\sub{upd} peak,
 and because the PZTC falls in the 
H\sub{upd} region, this consequentially shifts the PZTC.  This explains why the
PZTC of Pt is experimentally found to have nearly no shift with pH for the 111 surface (
-1~mV/pH unit\cite{Climent1997}), and a shift of 12~mV/pH unit for the 110 surface\cite{Frumkin1975}.

\begin{figure}
\includegraphics[width=\columnwidth]{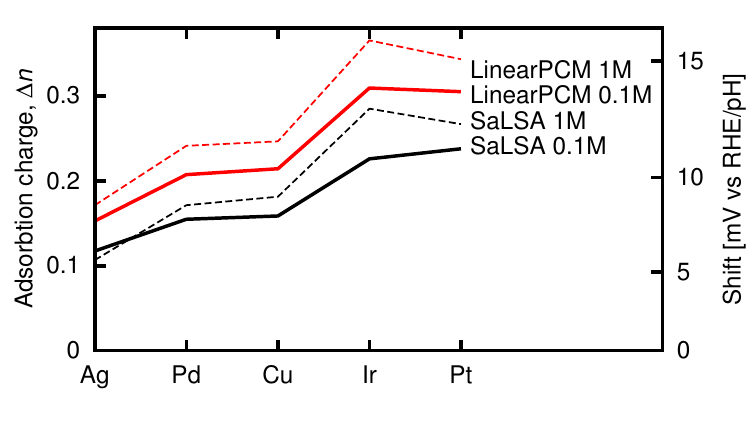}
\caption{Change in number of electrons, $n$, for H\sub{2}O adsorption
atop pristine 110 metal step edge sites, and corresponding predicted anomalous pH shifts.
\label{fig:OtherMetals}}
\end{figure}

The partially oxidized water that influences many reactions on the
platinum step edges is not unique to platinum.  We find that water
partially oxidizes on step edges more generally, and Figure~\ref{fig:OtherMetals}
reports the change in charge upon water adsorption on pristine 110 surface step edges
of metals frequently employed as catalysts.  The charge on the water changes somewhat
depending on the metal surface, but the phenomenon of partial oxidation is observed for 
all of these metal step edge sites.

\section*{Conclusions}

We identify the oxidative adsorption of water on step edges 
in platinum nanoparticles and stepped surfaces as the cause
of the anomalous p­­­­­­­­H effects observed for a number of processes
including the potential of zero charge and the onset of H\sub{upd}.
Although this does not rule out other phenomena such as electrolyte adsorption
or water dissociation, it indicates that they are not the cause for these anomalous pH effects.
Additionally, we find that the ionic strength is an important parameter, and we find 
good agreement between our DFT calculations and experimental results when the ionic strength
matches that used in experiment.  Our results predict that increasing the ionic strength will increase the 
pH effect for the onset of H\sub{upd} for the step edges.

These findings, which are relevant for a large potential and pH range,
and for a wide range of catalytic surface compositions, provide the framework for understanding the slowdown of reaction rates
(such as HER/HOR) in alkaline media, which we will investigate further in a subsequent paper.

Computationally, these results illustrate the importance of
directly calculating electrosorption valencies with fixed-potential DFT methods.
These electrosorption valencies can change as a function of potential,
leading to large energy differences between extrapolated values from
fixed-charge calculations, and those calculated directly from
fixed-potential calculations, as we show for the case of
water oxidatively adsorbed to step-edges.

Finally, these findings provide an improved way to find
the integrated area of a catalyst using the H\sub{upd} peaks.
Rather than assuming one electron transfer per proton, we
suggest a weighting factor of $\approx 1.2$~electrons
 per proton for the platinum step edge protons 
for more accurate integrated area estimates.

\section*{Acknowledgments}

KAS thanks T.P. Moffat for helpful discussions.
RS was supported by the Joint Center of Artificial Photosynthesis,
a DOE Energy Innovation Hub, supported through the Office of Science
of the U.S. Department of Energy under Award Number DE-SC0004993.

\footnotesize{
\providecommand*{\mcitethebibliography}{\thebibliography}
\csname @ifundefined\endcsname{endmcitethebibliography}
{\let\endmcitethebibliography\endthebibliography}{}

}

\end{document}